\documentclass[english,journal=jacsat,manuscript=article]{achemso}
\usepackage[T1]{fontenc}
\usepackage[utf8]{inputenc}
\usepackage{color}
\usepackage{babel}
\usepackage{amsmath}
\usepackage{amsthm}
\usepackage{graphicx}
\usepackage[unicode=true,pdfusetitle,
 bookmarks=true,bookmarksnumbered=false,bookmarksopen=false,
 breaklinks=false,pdfborder={0 0 0},pdfborderstyle={},backref=false,colorlinks=true]
 {hyperref}
\hypersetup{
 allcolors=blue}

\makeatletter

\title[Magnification effects in STM]{Magnification effects in scanning tunneling microscopy: the role
of surface radicals}
\author{Ruslan Zhachuk}
\affiliation[University of Aveiro]{Department of Physics \& I3N, University of Aveiro, Campus Santiago,
3810-193 Aveiro, Portugal}
\alsoaffiliation[Institute of Semiconductor Physics]{Institute of Semiconductor Physics, pr. Lavrentyeva 13, Novosibirsk
630090, Russia}
\email{zhachuk@gmail.com}
\author{José Coutinho}
\affiliation[University of Aveiro]{Department of Physics \& I3N, University of Aveiro, Campus Santiago,
3810-193 Aveiro, Portugal}
\email{jose.coutinho@ua.pt}
\phone{+351 234 247051}
\fax{+351 234 378197}

\setkeys{acs}{articletitle=true}

\makeatother

\begin{document}
\begin{abstract}
Scanning tunneling microscopy (STM) is a fundamental tool for determination
of the surface atomic structure. However, the interpretation of high
resolution microscopy images is not straightforward. In this paper
we provide a physical insight on how STM images can suggest atomic
locations which are distinctively different from the real ones. This
effect should be taken into account when interpreting high-resolution
STM images obtained on surfaces with directional bonds. It is shown
that spurious images are formed in the presence of polarized surface
radicals showing a pronounced angle with respect to the surface normal.
This issue has been overlooked within the surface science community
and often disregarded by experimentalists working with STM. Without
loss of generality, we illustrate this effect by the magnification
observed for pentamer-like structures on $(110)$, $(113)$ and $(331)$
surfaces of silicon and germanium.

\begin{figure}[H]

\includegraphics{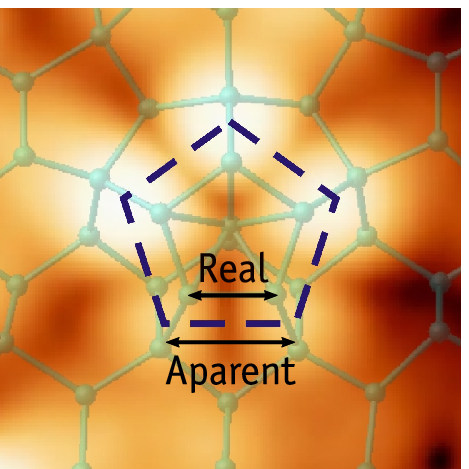}

\emph{TOC Graphical entry}
\end{figure}
\end{abstract}
The study of the atomic structure of a crystal surface often starts
with scanning tunneling microscopy (STM) measurements. The positions
of bright spots in high resolution STM images are then associated
with exact atomic coordinates on the surface under scrutiny. Although
being generally accepted, such a connection is known to be imperfect
with several deceiving cases being known. The STM images obtained
from Bi-terminated Si$(111)$ and Ge$(111)$ surfaces are clear examples\cite{mys10,zha11}
— while bright spots are observed atop of the Bi trimers at negative
bias and at large positive bias, at low positive bias the spots are
observed between trimers, where no atoms exist.

It is clear that the interpretation of high-resolution STM images
is not straightforward. According to a simplified view, STM images
obtained in a constant-current mode represent a combination of surface
topography and local density of electronic states.\cite{ter85,hof03}
Consequently, a full reasoning of such images, may only be possible
upon some knowledge of the surface atomic structure (which is often
unavailable), or with the assistance by \emph{ab-initio} electronic
structure calculations. 

While the connection between the surface atomic positions and bright
spots in STM images is often factual and correct (and so it may be
the resulting model), we show below that surface geometry misassignments
are likely to occur when the electronic polarization and hybridization
of states at surface atoms are overlooked. We illustrate the occurrence
of large offsets between bright spots from STM images and the corresponding
atomic nuclei with help of a paradigmatic example, namely with the
magnification of pentamers as seen in STM images of Si$(331)$.\cite{zha17}
Analogous pentameric structures were found on $(110)$ and $(113)$
surfaces of silicon and the $(110)$ surface of germanium.\cite{dab95,an00,ste04}

\begin{figure}
\includegraphics[clip,width=8cm]{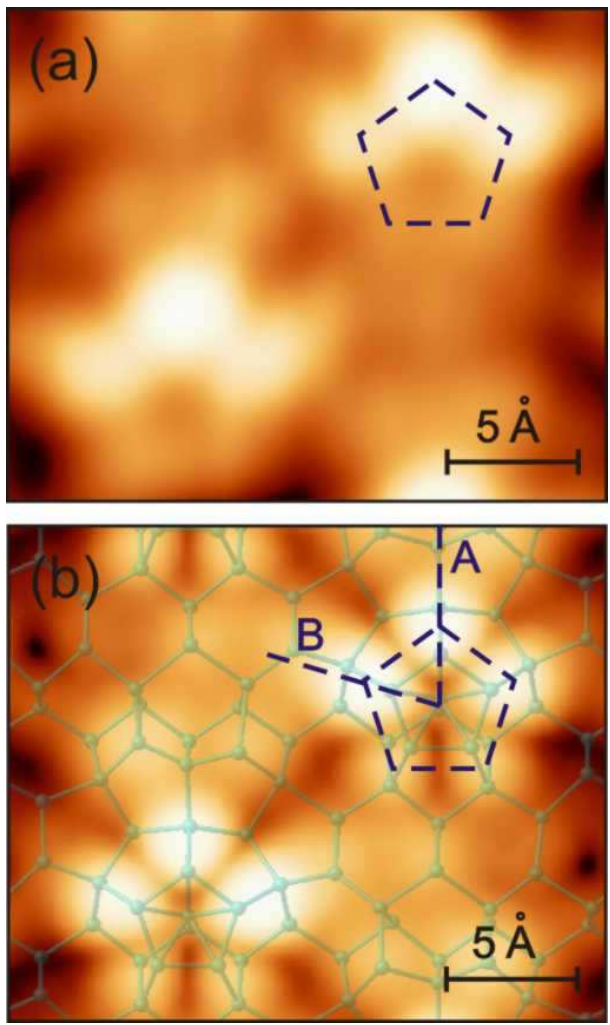}

\caption{\label{fig1}(a) Experimental high resolution STM image of the Si$(331)\textrm{-}12\times1$
surface. $U=+0.8$~V, $I=0.024$~nA. (b) Calculated STM image of
the Si$(331)\textrm{-}12\times1$ surface within the 8P-model of Ref.~\citenum{zha17}
(atomic positions are overlaid) using a voltage that corresponds to
states up to $+0.8$~eV above the theoretical Fermi level. Dashed
pentagons in (a) and (b) illustrate the apparent size of pentamer
features as derived from the center of bright spots. Dashed lines
marked by ‘A’ and ‘B’ are supporting directions used in the text for
discussion. Vertical and horizontal edges of the images are along
$[\bar{1}\bar{1}6]$ and $[\bar{1}10]$ crystallographic directions,
respectively.}
\end{figure}

Figure~\ref{fig1}(a) depicts an experimental high-resolution STM
image at positive bias ($U=+0.8$~V) of the Si$(331)\textrm{-}12\times1$
surface showing bright pentameric features.\cite{zha17} The bright
spots at the corners of one of the pentamers are connected by dashed
lines. The distance between two nearest spots is about $3.5\,\mathrm{\mathring{A}}$
as derived from the experimental STM image in Fig.~\ref{fig1}(a).

Figure~\ref{fig1}(b) shows the 8P atomic model for the Si$(331)\textrm{-}12\times1$
reconstruction,\cite{zha17} along with its respective STM image obtained
from the calculated LDOS at positive bias (considering states up to
$0.8$~eV above the Fermi level) within the Tersoff-Hamann approximation.\cite{ter85}
The image in Fig.~\ref{fig1}(b) represents the height at constant
integrated LDOS between +0.8V and Fermi level for Si$(331)$. It gives
an idea of how an STM image of the Si$(331)$ surface would look like
for an infinitesimally sharp tip, \emph{i.e.} close to a single atom
with an \emph{s}-orbital. The distance between the nearest bright
spots within the pentamers of Fig.~\ref{fig1}(b) agrees well the
analogous distance as derived from the experimental STM image of Fig.~\ref{fig1}(a).
However, the overlaid atomic structure shown in Fig.~\ref{fig1}(b)
clearly shows that the actual distance between nearest Si atoms in
each pentamer is $2.3\,\mathrm{\mathring{A}}$, corresponding to an
experimental deviation of more than 50\% with respect to atomic positions.
Due to its magnitude, this mismatch sustained an argument against
atomic models of $(110)$ and $(331)$ silicon surfaces involving
five-fold rings of Si atoms.\cite{sak09,tey17,zha17a}

\begin{figure}
\includegraphics[clip,width=8cm]{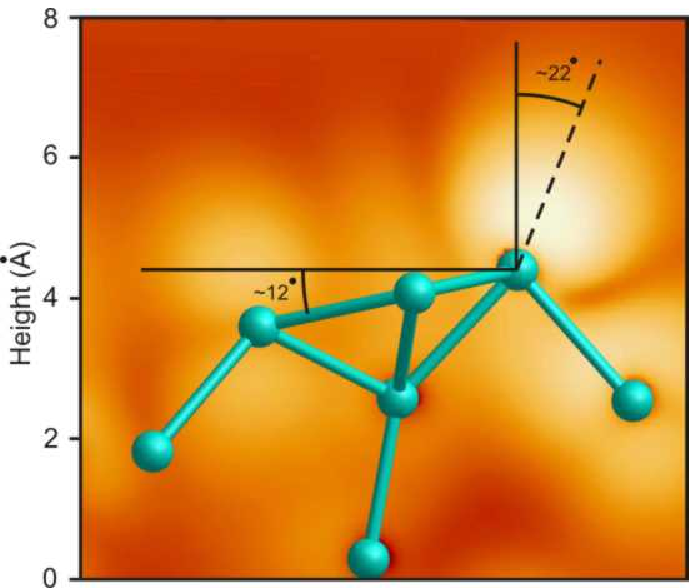}

\caption{\label{fig2}Vertical cut of the LDOS along the ‘A’ dashed line of
Fig.~\ref{fig1}(b), integrated over a $0.8$~eV energy window above
the calculated Fermi level. The intensity (brightness) is represented
on a logarithmic scale. The projection of a pentamer atomic structure
is overlaid. Solid black lines show horizontal and vertical directions
with respect to the surface. The dashed line is drawn from the apex
atom along the direction of maximum intensity of nearest bright spot.
Vertical and horizontal edges of the image are along $[331]$ and
$[\bar{1}\bar{1}6]$ crystallographic directions, respectively.}
\end{figure}

Figure~\ref{fig2} shows a \emph{vertical} cut of the height at constant
integrated LDOS (between +0.8V and Fermi level) across the ‘A’ dashed
line of Fig.~\ref{fig1}(b) ($[\bar{1}\bar{1}6]$ crystallographic
direction) combined with a projection of a pentamer atomic structure.
The brightest spot at the upper right corner of Fig.~\ref{fig2}
shows a high intensity for the empty LDOS located near the Si radical
at the upmost apex of the pentamer in Fig.~\ref{fig1}(b). Clearly,
the radical state does not point upwards (along $[331]$), rather
making an angle of about 22$^{\circ}$ with respect to surface normal
and away from the center of the pentamer. Since STM is intrinsically
sensitive to LDOS,\cite{ter85} and becasue the tip usually hoovers
between 4 to 10~Å above the surface,\cite{hof03} it becomes evident
that the \emph{slanted} radicals will project a magnified image of
the underlying pentamer. This explains the apparent contradiction
between the size of the pentamers from the atomistic model and those
derived directly from STM images (see Refs.~\citenum{sak09} and
\citenum{tey17}).

The distortion of the pentamers as observed by STM should not be confused
with another spurious effect which arises from the finite size of
the STM tip. The measured surface topography can be represented as
a convolution of the real surface topography and the shape of STM
tip.\cite{sch94} This effect leads to broadening of the objects protruded
out of the surface for the case of a dull tip. Comparing experimental
and calculated STM images in Fig.~\ref{fig1}, we realize that the
image in Fig.~\ref{fig1}(a) is indeed affected by the latter effect
– the experimental image is more diffuse than the calculated one.
The contribution of the finite tip size effect to the magnification
of the pentamers must be however minor. A 50\% increase of the pentamer
size (as observed in Fig.~\ref{fig1}(b)) would imply a severe and
inhomogeneous broadening of the bright spots, which would not fit
the experimental data.

\begin{figure}
\includegraphics[clip,width=8cm]{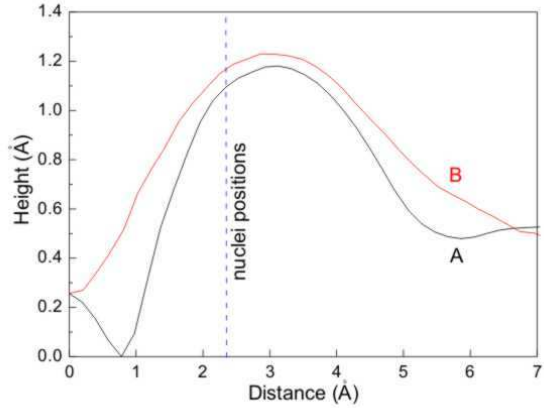}

\caption{\label{fig3}Height profiles along dashed lines ‘A’ and ‘B’ in Fig.~\ref{fig1}(b)
with origin at the center of a pentamer. The position of the atomic
nuclei at the pentamer vertices is indicated by a dashed line.}
\end{figure}

Figure~\ref{fig3} shows the height profiles measured across the
bright spots in the calculated STM image of Fig.~\ref{fig1}(b).
The horizontal axis runs from the center of a pentamer either along
‘A’ or ‘B’ dashed lines. The vertical dashed line in Fig.~\ref{fig3}
represents the actual position of the Si nuclei in the pentamer, clearly
showing that they do not coincide with positions of the maxima. The
lateral bright spot along ‘B’ is more diffuse then the spot close
to the central apex along ‘A’. This feature is also observed on the
experimental STM images, and that is due to the inclination of a pentamer
plane with respect to the surface normal by about 22$^{\circ}$ towards
the $[11\bar{6}]$ direction (see Fig.~\ref{fig2}).

The observation of \emph{spurious images by} STM as described above
is not limited to the case of pentamers in the $(110)$, $(113)$,
$(331)$ surfaces of silicon and germanium. It should be observed
virtually on any surface having directional dangling bonds whose axes
are not normal to the surface. Of course this will essentially depend
on the specific surface bonding and reconstruction details, which
can be scrutinized by first-principles atomistic methods.

Establishing a direct connection between the positions of bright spots
in STM images and the positions of atomic nuclei can lead to considerable
sizing and interpretation errors. For example, the Si surface orientation
composed of regularly spaced $(111)$ terraces and triple steps has
been erroneously identified. While an early STM study on this subject
indicated a $(557)$ orientation {[}9.5$^{\circ}$ off the $(111)$
plane{]},\cite{kir01} latter works including high resolution diffraction
data report a $(7\,7\,10)$ orientation {[}10.0$^{\circ}$ off the
$(111)$ plane{]}.\cite{tey06,zha09,zha14,pre12,leo16,leo17}

In conclusion, we demonstrated that the positions of bright spots
in high resolution STM images and the actual coordinates of atomic
nuclei may differ substantially. Elusive images can be formed on surfaces
having dangling-bond states whose main axis differs from the surface
normal, and that may lead to misinterpretation of the experimental
data. Pentameric structures on Si(331) observed by STM provided a
showcase for the above effect, which if taken into account, reconciles
the conflicting theoretical and measured sizes of pentamers occurring
on $(110)$, $(113)$ and $(331)$ surfaces of Si and Ge.

\section*{Experimental and computational methods}

The measurements were performed in an ultrahigh vacuum chamber ($7\times10^{-11}\,\mathrm{Torr}$)
on a system equipped with an Omicron STM. A clean Si(331) surface
was prepared by sample flash annealing at 1250$^{\circ}$C for 1 min
followed by stepwise cooling with 2$^{\circ}$C per minute steps within
a temperature range 400-850~$^{\circ}\mathrm{C}$. The STM images
were recorded at room temperature in the constant-current mode using
an electrochemically etched tungsten tip.

Electronic structure calculations were carried using the density-functional
plane-wave $\mathtt{VASP}$ code.\cite{kre93,kre94,kre96,kre96a}
The many-body exchange-correlation interactions were accounted for
within the semi-local generalized gradient approximation.\cite{per96}
Core electrons were replaced by efficient projector-augmented wave
(PAW) potentials,\cite{blo94,kre99} allowing a description of the
Kohn-Sham valence states using plane-waves with a cut-off energy of
250~eV.

The Si$(331)\textrm{-}12\times1$ surface was represented by a periodic
slab, consisting of ten atomic bilayers of silicon and $25\,\mathrm{\mathring{A}}$
of vacuum along the surface normal. The bottom side of the slab was
passivated by hydrogen atoms, while the top side was prepared according
to the 8P-model of the $12\times1$ reconstruction as reported in
Ref.~\citenum{zha17}. Atomic coordinates of the topmost eight bilayers
were fully relaxed until the largest Hellmann-Feynman force became
smaller than $0.015\,\mathrm{eV/\mathring{A}}$. The electron density
(and potential terms) was obtained by sampling the band structure
within the Brillouin zone using a $4\times4\times1$ $\mathbf{k}$-point
grid.\cite{mon76} The local density of states (LDOS) for electrons
at the surface was obtained from the Kohn-Sham eigenvalues and eigenfunctions.
The STM images were calculated within the Tersoff-Hamann approach,\cite{ter85}
assuming a constant current mode. The $\mathtt{WSXM}$ software was
used to process the experimental and calculated STM images.\cite{hor07}
\begin{acknowledgement}
The authors thank S. Teys for the given STM images. This work was
funded by the Russian Foundation for Basic Research (Project No. 18-02-00025),
the Fundação para a Ciência e a Tecnologia (FCT) under the contract
UID/CTM/50025/2013, and by FEDER funds through the COMPETE 2020 Program.
\end{acknowledgement}
%\bibliography{refs}

\begin{mcitethebibliography}{29}
\providecommand*\natexlab[1]{#1}
\providecommand*\mciteSetBstSublistMode[1]{}
\providecommand*\mciteSetBstMaxWidthForm[2]{}
\providecommand*\mciteBstWouldAddEndPuncttrue
  {\def\EndOfBibitem{\unskip.}}
\providecommand*\mciteBstWouldAddEndPunctfalse
  {\let\EndOfBibitem\relax}
\providecommand*\mciteSetBstMidEndSepPunct[3]{}
\providecommand*\mciteSetBstSublistLabelBeginEnd[3]{}
\providecommand*\EndOfBibitem{}
\mciteSetBstSublistMode{f}
\mciteSetBstMaxWidthForm{subitem}{(\alph{mcitesubitemcount})}
\mciteSetBstSublistLabelBeginEnd
  {\mcitemaxwidthsubitemform\space}
  {\relax}
  {\relax}

\bibitem[Myslive{\v c}ek \latin{et~al.}(2010)Myslive{\v c}ek, Dvo{\v r}{\' a}k,
  Str{\' o}{\. z}ecka, and Voigtl{\"a}nder]{mys10}
Myslive{\v c}ek,~J.; Dvo{\v r}{\' a}k,~F.; Str{\' o}{\. z}ecka,~A.;
  Voigtl{\"a}nder,~B. Scanning tunneling microscopy contrast in lateral {Ge-Si}
  nanostructures on {Si(111)}-$\sqrt{3}\times\sqrt{3}$-{Bi}. \emph{Physical
  Review B} \textbf{2010}, \emph{81}, 245427\relax
\mciteBstWouldAddEndPuncttrue
\mciteSetBstMidEndSepPunct{\mcitedefaultmidpunct}
{\mcitedefaultendpunct}{\mcitedefaultseppunct}\relax
\EndOfBibitem
\bibitem[Zhachuk and Coutinho(2011)Zhachuk, and Coutinho]{zha11}
Zhachuk,~R.; Coutinho,~J. Ab initio study of height contrast in scanning
  tunneling microscopy of {Ge/Si} surface layers grown on {Si}(111) in presence
  of {Bi}. \emph{Physical Review B} \textbf{2011}, \emph{84}, 193405\relax
\mciteBstWouldAddEndPuncttrue
\mciteSetBstMidEndSepPunct{\mcitedefaultmidpunct}
{\mcitedefaultendpunct}{\mcitedefaultseppunct}\relax
\EndOfBibitem
\bibitem[Tersoff and Hamann(1985)Tersoff, and Hamann]{ter85}
Tersoff,~J.; Hamann,~D.~R. Theory of the scanning tunneling microscope.
  \emph{Physical Review B} \textbf{1985}, \emph{31}, 805\relax
\mciteBstWouldAddEndPuncttrue
\mciteSetBstMidEndSepPunct{\mcitedefaultmidpunct}
{\mcitedefaultendpunct}{\mcitedefaultseppunct}\relax
\EndOfBibitem
\bibitem[Hofer(2003)]{hof03}
Hofer,~W.~A. Challenges and errors: interpreting high resolution images in
  scanning tunneling microscopy. \emph{Progress in Surface Science}
  \textbf{2003}, \emph{71}, 147\relax
\mciteBstWouldAddEndPuncttrue
\mciteSetBstMidEndSepPunct{\mcitedefaultmidpunct}
{\mcitedefaultendpunct}{\mcitedefaultseppunct}\relax
\EndOfBibitem
\bibitem[Zhachuk and Teys(2017)Zhachuk, and Teys]{zha17}
Zhachuk,~R.; Teys,~S. Pentagons in the {Si}(331)-($12\times1$) surface
  reconstruction. \emph{Physical Review B} \textbf{2017}, \emph{95},
  041412\relax
\mciteBstWouldAddEndPuncttrue
\mciteSetBstMidEndSepPunct{\mcitedefaultmidpunct}
{\mcitedefaultendpunct}{\mcitedefaultseppunct}\relax
\EndOfBibitem
\bibitem[D{\k a}browski \latin{et~al.}(1995)D{\k a}browski, M{\"u}ssig, and
  Wolff]{dab95}
D{\k a}browski,~J.; M{\"u}ssig,~H.-J.; Wolff,~G. A novel surface
  reconstruction: subsurface interstitials stabilize {Si(113)}-$3\times2$.
  \emph{Surface Science} \textbf{1995}, \emph{331-333}, 1022\relax
\mciteBstWouldAddEndPuncttrue
\mciteSetBstMidEndSepPunct{\mcitedefaultmidpunct}
{\mcitedefaultendpunct}{\mcitedefaultseppunct}\relax
\EndOfBibitem
\bibitem[An \latin{et~al.}(2000)An, Yoshimura, Ono, and Ueda]{an00}
An,~T.; Yoshimura,~M.; Ono,~I.; Ueda,~K. Elemental structure in
  {Si(110)}-"$16\times2$" revealed by scanning tunneling microscopy.
  \emph{Physical Review B} \textbf{2000}, \emph{61}, 3006\relax
\mciteBstWouldAddEndPuncttrue
\mciteSetBstMidEndSepPunct{\mcitedefaultmidpunct}
{\mcitedefaultendpunct}{\mcitedefaultseppunct}\relax
\EndOfBibitem
\bibitem[Stekolnikov \latin{et~al.}(2004)Stekolnikov, Furthm{\"u}ller, and
  Bechstedt]{ste04}
Stekolnikov,~A.~A.; Furthm{\"u}ller,~J.; Bechstedt,~F. Structural elements on
  reconstructed {Si} and {Ge}(110) surfaces. \emph{Physical Review B}
  \textbf{2004}, \emph{70}, 045305\relax
\mciteBstWouldAddEndPuncttrue
\mciteSetBstMidEndSepPunct{\mcitedefaultmidpunct}
{\mcitedefaultendpunct}{\mcitedefaultseppunct}\relax
\EndOfBibitem
\bibitem[Sakamoto \latin{et~al.}(2009)Sakamoto, Setvin, Mawatari, Eriksson,
  Miki, and Uhrberg]{sak09}
Sakamoto,~K.; Setvin,~M.; Mawatari,~K.; Eriksson,~P. E.~J.; Miki,~K.;
  Uhrberg,~R. I.~G. Electronic structure of the {Si}(110)-($16\times2$)
  surface: high-resolution {ARPES} and {STM} investigation. \emph{Physical
  Review B} \textbf{2009}, \emph{79}, 045304\relax
\mciteBstWouldAddEndPuncttrue
\mciteSetBstMidEndSepPunct{\mcitedefaultmidpunct}
{\mcitedefaultendpunct}{\mcitedefaultseppunct}\relax
\EndOfBibitem
\bibitem[Teys(2017)]{tey17}
Teys,~S.~A. Different {STM} images of the superstructure on a clean
  {Si}(133)-$6\times2$ surface. \emph{JETP Letters} \textbf{2017}, \emph{105},
  477\relax
\mciteBstWouldAddEndPuncttrue
\mciteSetBstMidEndSepPunct{\mcitedefaultmidpunct}
{\mcitedefaultendpunct}{\mcitedefaultseppunct}\relax
\EndOfBibitem
\bibitem[Zhachuk and Coutinho(2017)Zhachuk, and Coutinho]{zha17a}
Zhachuk,~R.; Coutinho,~J. Comment on "{Different} {STM} images of the
  superstructure on a clean {Si}(133)-$6\times2$ surface" ({JETP Letters} 105,
  477 (2017)). \emph{JETP Letters} \textbf{2017}, \emph{106}, 346\relax
\mciteBstWouldAddEndPuncttrue
\mciteSetBstMidEndSepPunct{\mcitedefaultmidpunct}
{\mcitedefaultendpunct}{\mcitedefaultseppunct}\relax
\EndOfBibitem
\bibitem[Schwarz \latin{et~al.}(1994)Schwarz, Haefke, Reimann, and
  G{\"u}ntherodt]{sch94}
Schwarz,~U.~D.; Haefke,~H.; Reimann,~P.; G{\"u}ntherodt,~H.-J. Tip artefacts in
  scanning force microscopy. \emph{Journal of Microscopy} \textbf{1994},
  \emph{173}, 183\relax
\mciteBstWouldAddEndPuncttrue
\mciteSetBstMidEndSepPunct{\mcitedefaultmidpunct}
{\mcitedefaultendpunct}{\mcitedefaultseppunct}\relax
\EndOfBibitem
\bibitem[Kirakosian \latin{et~al.}(2001)Kirakosian, Bennewitz, Crain, Fauster,
  Lin, Petrovykh, and Himpsel]{kir01}
Kirakosian,~A.; Bennewitz,~R.; Crain,~J.~N.; Fauster,~T.; Lin,~J.-L.;
  Petrovykh,~D.~Y.; Himpsel,~F.~J. Atomically accurate {Si} grating with
  5.73~nm period. \emph{Applied Physics Letters} \textbf{2001}, \emph{79},
  1608\relax
\mciteBstWouldAddEndPuncttrue
\mciteSetBstMidEndSepPunct{\mcitedefaultmidpunct}
{\mcitedefaultendpunct}{\mcitedefaultseppunct}\relax
\EndOfBibitem
\bibitem[Teys \latin{et~al.}(2006)Teys, Romanyuk, Zhachuk, and
  Olshanetsky]{tey06}
Teys,~S.~A.; Romanyuk,~K.~N.; Zhachuk,~R.~A.; Olshanetsky,~B.~Z. Orientation
  and structure of triple step staircase on vicinal {Si}(111) surfaces.
  \emph{Surface Science} \textbf{2006}, \emph{600}, 4878\relax
\mciteBstWouldAddEndPuncttrue
\mciteSetBstMidEndSepPunct{\mcitedefaultmidpunct}
{\mcitedefaultendpunct}{\mcitedefaultseppunct}\relax
\EndOfBibitem
\bibitem[Zhachuk and Pereira(2009)Zhachuk, and Pereira]{zha09}
Zhachuk,~R.; Pereira,~S. Comment on "{A}tomic structure model of the
  reconstructed {Si}(557) surface with a triple step structure: adatom-parallel
  dimer model". \emph{Physical Review B} \textbf{2009}, \emph{79}, 077401\relax
\mciteBstWouldAddEndPuncttrue
\mciteSetBstMidEndSepPunct{\mcitedefaultmidpunct}
{\mcitedefaultendpunct}{\mcitedefaultseppunct}\relax
\EndOfBibitem
\bibitem[Zhachuk \latin{et~al.}(2014)Zhachuk, Teys, Coutinho, Rayson, and
  Briddon]{zha14}
Zhachuk,~R.; Teys,~S.; Coutinho,~J.; Rayson,~M.~J.; Briddon,~P.~R. Static and
  dynamic buckling of reconstructions at triple steps on {Si}(111) surfaces.
  \emph{Applied Physics Letters} \textbf{2014}, \emph{105}, 171602\relax
\mciteBstWouldAddEndPuncttrue
\mciteSetBstMidEndSepPunct{\mcitedefaultmidpunct}
{\mcitedefaultendpunct}{\mcitedefaultseppunct}\relax
\EndOfBibitem
\bibitem[Pr{\'e}vot \latin{et~al.}(2012)Pr{\'e}vot, Leroy, Croset, Garreau,
  Coati, and M{\"u}ller]{pre12}
Pr{\'e}vot,~G.; Leroy,~F.; Croset,~B.; Garreau,~Y.; Coati,~A.; M{\"u}ller,~P.
  Step-induced elastic relaxation and surface structure of the {Si}(7710)
  surface. \emph{Surface Science} \textbf{2012}, \emph{606}, 209\relax
\mciteBstWouldAddEndPuncttrue
\mciteSetBstMidEndSepPunct{\mcitedefaultmidpunct}
{\mcitedefaultendpunct}{\mcitedefaultseppunct}\relax
\EndOfBibitem
\bibitem[Le{\'o}n \latin{et~al.}(2016)Le{\'o}n, Drees, Wippermann, Marz, and
  Hoffmann-Vogel]{leo16}
Le{\'o}n,~C.~P.; Drees,~H.; Wippermann,~S.~M.; Marz,~M.; Hoffmann-Vogel,~R.
  Atomic-scale imaging of the surface dipole distribution of stepped surfaces.
  \emph{Journal of Physical Chemistry Letters} \textbf{2016}, \emph{7},
  426\relax
\mciteBstWouldAddEndPuncttrue
\mciteSetBstMidEndSepPunct{\mcitedefaultmidpunct}
{\mcitedefaultendpunct}{\mcitedefaultseppunct}\relax
\EndOfBibitem
\bibitem[Le{\'o}n \latin{et~al.}(2017)Le{\'o}n, Drees, Wippermann, Marz, and
  Hoffmann-Vogel]{leo17}
Le{\'o}n,~C.~P.; Drees,~H.; Wippermann,~S.~M.; Marz,~M.; Hoffmann-Vogel,~R.
  Atomically resolved scanning force studies of vicinal {Si}(111).
  \emph{Physical Review B} \textbf{2017}, \emph{95}, 245412\relax
\mciteBstWouldAddEndPuncttrue
\mciteSetBstMidEndSepPunct{\mcitedefaultmidpunct}
{\mcitedefaultendpunct}{\mcitedefaultseppunct}\relax
\EndOfBibitem
\bibitem[Kresse and Hafner(1993)Kresse, and Hafner]{kre93}
Kresse,~G.; Hafner,~J. Ab initio molecular dynamics for liquid metals.
  \emph{Physical Review B} \textbf{1993}, \emph{47}, 558\relax
\mciteBstWouldAddEndPuncttrue
\mciteSetBstMidEndSepPunct{\mcitedefaultmidpunct}
{\mcitedefaultendpunct}{\mcitedefaultseppunct}\relax
\EndOfBibitem
\bibitem[Kresse and Hafner(1994)Kresse, and Hafner]{kre94}
Kresse,~G.; Hafner,~J. Ab initio molecular-dynamics simulation of the
  liquid-metal-amorphous-semiconductor transition in germanium. \emph{Physical
  Review B} \textbf{1994}, \emph{49}, 14251\relax
\mciteBstWouldAddEndPuncttrue
\mciteSetBstMidEndSepPunct{\mcitedefaultmidpunct}
{\mcitedefaultendpunct}{\mcitedefaultseppunct}\relax
\EndOfBibitem
\bibitem[Kresse and Furthm{\"u}ller(1996)Kresse, and Furthm{\"u}ller]{kre96}
Kresse,~G.; Furthm{\"u}ller,~J. Efficient iterative schemes for ab initio
  total-energy calculations using a plane-wave basis set. \emph{Physical Review
  B} \textbf{1996}, \emph{54}, 11169\relax
\mciteBstWouldAddEndPuncttrue
\mciteSetBstMidEndSepPunct{\mcitedefaultmidpunct}
{\mcitedefaultendpunct}{\mcitedefaultseppunct}\relax
\EndOfBibitem
\bibitem[Kresse and Furthm{\"u}ller(1996)Kresse, and Furthm{\"u}ller]{kre96a}
Kresse,~G.; Furthm{\"u}ller,~J. Efficiency of ab-initio total energy
  calculations for metals and semiconductors using a plane-wave basis set.
  \emph{Computational Materials Science} \textbf{1996}, \emph{6}, 15\relax
\mciteBstWouldAddEndPuncttrue
\mciteSetBstMidEndSepPunct{\mcitedefaultmidpunct}
{\mcitedefaultendpunct}{\mcitedefaultseppunct}\relax
\EndOfBibitem
\bibitem[Perdew \latin{et~al.}(1996)Perdew, Burke, and Ernzerhof]{per96}
Perdew,~J.~P.; Burke,~K.; Ernzerhof,~M. Generalized {G}radient {A}pproximation
  made simple. \emph{Physical Review Letters} \textbf{1996}, \emph{77},
  3865\relax
\mciteBstWouldAddEndPuncttrue
\mciteSetBstMidEndSepPunct{\mcitedefaultmidpunct}
{\mcitedefaultendpunct}{\mcitedefaultseppunct}\relax
\EndOfBibitem
\bibitem[Bl{\"o}chl(1994)]{blo94}
Bl{\"o}chl,~P.~E. Projector augmented-wave method. \emph{Physical Review B}
  \textbf{1994}, \emph{50}, 17953\relax
\mciteBstWouldAddEndPuncttrue
\mciteSetBstMidEndSepPunct{\mcitedefaultmidpunct}
{\mcitedefaultendpunct}{\mcitedefaultseppunct}\relax
\EndOfBibitem
\bibitem[Kresse and Joubert(1999)Kresse, and Joubert]{kre99}
Kresse,~G.; Joubert,~D. From ultrasoft pseudopotentials to the projector
  augmented-wave method. \emph{Physical Review B} \textbf{1999}, \emph{59},
  1758\relax
\mciteBstWouldAddEndPuncttrue
\mciteSetBstMidEndSepPunct{\mcitedefaultmidpunct}
{\mcitedefaultendpunct}{\mcitedefaultseppunct}\relax
\EndOfBibitem
\bibitem[Monkhorst and Pack(1976)Monkhorst, and Pack]{mon76}
Monkhorst,~H.~J.; Pack,~J.~D. Special points for {B}rillouin-zone integrations.
  \emph{Physical Review B} \textbf{1976}, \emph{13}, 5188\relax
\mciteBstWouldAddEndPuncttrue
\mciteSetBstMidEndSepPunct{\mcitedefaultmidpunct}
{\mcitedefaultendpunct}{\mcitedefaultseppunct}\relax
\EndOfBibitem
\bibitem[Horcas \latin{et~al.}(2007)Horcas, Fern{\'a}ndez,
  G{\'o}mez-Rodr{\'i}guez, Colchero, G{\'o}mez-Herrero, and Baro]{hor07}
Horcas,~I.; Fern{\'a}ndez,~R.; G{\'o}mez-Rodr{\'i}guez,~J.~M.; Colchero,~J.;
  G{\'o}mez-Herrero,~J.; Baro,~A.~M. {WSXM}: A software for scanning probe
  microscopy and a tool for nanotechnology. \emph{Review of Scientific
  Instruments} \textbf{2007}, \emph{78}, 013705\relax
\mciteBstWouldAddEndPuncttrue
\mciteSetBstMidEndSepPunct{\mcitedefaultmidpunct}
{\mcitedefaultendpunct}{\mcitedefaultseppunct}\relax
\EndOfBibitem
\end{mcitethebibliography}

\providecommand{\latin}[1]{#1}
\makeatletter
\providecommand{\doi}
  {\begingroup\let\do\@makeother\dospecials
  \catcode`\{=1 \catcode`\}=2\doi@aux}
\providecommand{\doi@aux}[1]{\endgroup\texttt{#1}}
\makeatother
\providecommand*\mcitethebibliography{\thebibliography}
\csname @ifundefined\endcsname{endmcitethebibliography}
  {\let\endmcitethebibliography\endthebibliography}{}

\end{document}